# A Comparative Study of Parametric Regression Models to Detect Breakpoint in Traffic Fundamental Diagram

Emmanuel Kidando
Department of Civil and Environmental Engineering
Cleveland State University
e.kidando@csuoio.edu

Angela E. Kitali (corresponding author)
Department of Civil and Environmental Engineering
Florida International University
akita002@fiu.edu

Boniphace Kutela, Ph.D.
Roadway Safety Program
Texas A&M Transportation Institute
b-kutela@tti.tamu.edu

Thobias Sando
School of Engineering
University of North Florida
t.sando@unf.edu

**ABSTRACT**

A speed threshold is a crucial parameter in breakdown and capacity distribution analysis as it defines the boundary between free-flow and congested regimes. However, literature on approaches to establish the breakpoint value for detecting breakdown events is limited. Most of existing studies rely on the use of either visual observation or predefined thresholds. These approaches may not be reliable considering the variations associated with field data. Thus, this study compared the performance of two data-driven methods, i.e., logistic function (LGF) and two-regime models, used to establish the breakpoint from traffic flow variables. The two models were calibrated using urban freeway traffic data. The models' performance results revealed that with less computation efforts, the LGF has slightly better prediction accuracy than the two-regime model. Although the two-regime model had relatively lower performance, it can be useful in identifying the transitional state.

**INTRODUCTION**

Traffic breakdown is a common problem that affects the stability and reliability of traffic operations, especially in urban areas. It causes a sudden drop in speed when demand exceeds capacity. The traffic breakdown phenomenon is known to be a stochastic (i.e. the traffic flow parameters exhibit random characteristics) process that can occur at any traffic flow level and any location of the highway (Kim et al. 2010; Dong and Mahmassani 2009). The probability distribution that governs the breakdown process is often used to estimate capacity (Dong and Mahmassani, 2009; Brilon et al., 2005; Persaud et al. 1998; Lorenz and Elefteriadou 2014). Accordingly, a better understanding of the breakdown process improves the success of the traffic management strategies in tackling traffic congestion (Brilon et al., 2005; Persaud et al. 1998; Lorenz and Elefteriadou 2014; Elefteriadou and Lertworawanich 2003; Matt and Elefteriadou 2001; Xu, et al. 2013).

Previous studies have extensively investigated the stochastic characteristics of the breakdown process that leads to highways reaching capacity. However, literature on approaches

3for establishing the breakpoint (threshold) value to detect breakdown event is sparse (Kidando et al. 2019). Many studies visually inspect the traffic fundamental diagram (i.e., the relationships between flow, speed, and density) to detect a breakpoint value in defining the breakdown event (Kim et al. 2010; Dong and Mahmassani 2009; Shao and Liu 2015; Dowling et al. 2008; Dehman 2013). This approach may not be reliable considering the variations associated with field data (Kidando et al. 2019; Dehman 2013). The variations in the field data are caused by the stochasticity of the microscopic traffic flow characteristics of individual vehicles (e.g., lane changing behaviors and individual vehicle speed), weather conditions, etc. Other studies have used a pre-defined value as the estimated breakpoint (Shao and Liu 2015; Jun and Lim 2009; Iqbal et al. 2017; Yeon et al. 2009). Nonetheless, accurate modeling and prediction of the breakdown process and the capacity probability distribution require the breakpoint value to be estimated using field data to capture the actual characteristics of the study corridor (Kidando et al. 2019; Dehman 2013). This will help to accurately estimate the breakpoint value in the analysis of capacity, breakdown process, and congestion modeling.

The single-regime models and multi-regime models are statistical approaches that can be used to calibrate the breakpoint value while accounting for the stochastic characteristics of the traffic flow. Among the single-regime traffic flow models, the logistic function (LGF) proposed by Wang et al. (2011) estimates the breakpoint value from empirical data. The use of this function to calibrate the speed-density relationship is among the latest traffic flow models. Literature indicates that the LGF can fit empirical data with reasonable accuracy (Wang et al. 2013). Furthermore, the LGF parameters have a physical meaning in traffic flow theory (Wang et al. 2011; Wang et al. 2013). Among the fitted parameters in the LGF, the breakpoint value that defines the boundary between free-flow and congested regime is estimated.

Multi-regime models provide an alternative way of determining the breakpoint in the speed-density relationship. Unlike the single-regime models, the multi-regime models use two or more curves separated by the breakpoint to calibrate different traffic regimes (Qu et al. 2015). Some of the multi-regime models that have been proposed in the literature include the Edie model, the two-regime model, the modified Greenberg model, and three-regime models (Drake et al. 1967; May 1990; Edie 1967). These models are criticized on how the breakpoints are identified in the analyses. The breakpoint separating traffic regimes in these models are estimated based on observational judgment (Wang et al. 2011; Wang et al. 2013). Recent advances in computation have made calibrating the breakpoint(s) in the multi-regime model possible (Kidando et al. 2019).

This study attempts to compare the single-regime and multi-regime models, based on the LGF and two-regime models respectively, in calibrating the speed breakpoint for identifying the breakdown event. The two models are compared in terms of their computation effort as well as their prediction accuracy. As it is well known that the traffic fundamental diagrams are stochastic in nature (Wang et al. 2009; Ni 2015; Lin et al. 2012; Sopasakis 2004), this study uses the Markov Chain Monte Carlo (MCMC) simulations to estimate the model parameters. The resulting estimates are distributions accounting for uncertainty rather than point estimates.

**LITERATURE REVIEW**

Since the breakdown process is stochastic, the challenge is how to properly define it. This is an important step that may ultimately affect the capacity of the road and its operational efficiency. The traffic flow speed is a measure that has mainly been used to identify breakdown whereby the breakpoint value is determined by a speed drop or a specific breakpoint (Kidando et al. 2019; Kondyli et al. 2013). Various approaches have been used to estimate the breakpoint value for defining the breakdown process. As indicated in Table 1, these approaches can be broadly grouped into three major categories: (i) visual observation, (ii) predefined breakpoints, and (iii) data-driven approaches.

A majority of the previous studies used a visual inspection of the speed-flow curve to extract the speed breakpoint focused on the drop in speeds (Kim et al. 2010; Dong and Mahmassani 2009; Sasahara and Elefteriadou 2019; Elefteriadou and Lertworawanich 2003; Shao and Liu 2015; Dowling et al. 2008; )( (Kim et al. 2010; Dong and Mahmassani 2009; Sasahara and Elefteriadou 2019; Elefteriadou and Lertworawanich 2003; Shao and Liu 2015; Dowling et al. 2008; Ma, et al. 2012; Modi et al. 2014; Laflamme and Ossenbruggen 2015; Ossenbruggen 2016) For instance, one of the most recent studies analyzed the impact of spillback occurrence on freeway off-ramps' likelihood of breakdown (Sasahara and Elefteriadou 2019. A speed drop of at least 10% of the free-flow speed was adopted as one of the criteria for defining the breakpoint. Another study combined two criteria to define the breakpoint: 8% speed drop and 75% of the free-flow speed (Asgharzadeh and Kondyli 2019).

Table 1. Summary of literature on defining breakpoint

| Method category | Reference | Corridor characteristics | Data source | Data type | Breakdown point estimation method | Breakpoint | Purpose of estimating breakdown |
|---|---|---|---|---|---|---|---|
| Visual observation | Elefteriadou and Lertworawanich (2003) | 2-highway 401 freeway segment in Canada | Loop detectors | 15-min speed and volume data | Time series plots for speed and flow | 56 mph | Estimating freeway capacity |
| | Dowling et al. (2008) | US-101 freeway section with an auxiliary lane plus a high-occupancy-vehicle California | Loop detectors | 5-min speed and flow | Plotting the change in speed between observations and identifying the speed at which speed is most unstable | 45 mph | Predict the impact of intelligent transportation systems on freeway queue discharge flow variability |
| | Ma et al. (2012) | Expressway diverge section Japan | Double-loop detectors | 5-min aggregated traffic flow rates and average speed | Flow rate-speed diagram | NA | Propose a lane-based approach to identify breakdown for diverge sections on express lane |
| | Dehman (2013) | Four three-lane on-ramp freeway segments in Wisconsin | Traffic detectors | 5-min occupancy, volume, and speed | Fundamental speed-flow diagram | NA | Examine the influence of the mix between ramp and mainline flows on breakdown and capacity characteristics |
| | Modi et al. (2014) | 22 basic-freeway segments Florida | RTMS | 1- and 5- minutes aggregated volume, speed, and occupancy data | Speed time series plots | NA | Estimate freeway capacity |
| | Laflamme and Ossenbruggen (2014) | I-93 freeway segment in Wisconsin | Side-fire radar device | 15-min aggregated vehicle counts, average speed, occupancy, and speed of individual vehicles | Speed functional data analysis traces | 48 mph | Evaluate the effect of volatility on the probability of highway breakdown |
| | Ossenbruggen (2016) | I-93 freeway segment in Wisconsin | Side-fire radar device | 15-min harmonic mean of speed and flow | Visual observation | 25 mph and 40 mph | Evaluate the probability of breakdown and recovery |
| Predefined breakpoints | Dong and Mahmassani (2009) | I-495 freeway segment in California | Loop detectors | 5-min average flow and speed | Speed flow curve | 10 mph | Develop and online prediction of travel time reliability based on real- |
| | | | | | | | world measurements in light of the probabilistic character |
| | Yeon et al. (2009) | 8-non-basic freeway segment in Philadelphia | RTMS | 5-min speed and volume | | 50 mph | Variation of capacity flows by day of the week and time of day |
| | Shiomi et al. (2011) | Expressway segment in Japan | Loop detectors | 5-min average speed | Fundamental diagram | 31 mph | Establish the relationship between breakdown probability and traffic flow rate |
| | Kondyli et al. (2013) | 5-freeway ramp merging segments in Minnesota, Canada, and California | Loop detectors | 1-min aggregated volume, speed, and occupancy | Speed-based/ occupancy-based and volume-occupancy correlation-based algorithms | 10 mph speed and 5% occupancy | Develop probabilistic model to predict traffic flow breakdown |
| Data-driven approaches | Shiomi (2016) | 3-lane expressway in Japan | Dual-loop detector | 5-min aggregated mean average speed and flow | Fundamental diagram | 34-43 mph | Propose a control scheme of lane traffic flow for managing the uncertainty in traffic breakdown caused by the unbalanced lane-use |
| | Xie et al. (2014) | 4-diverge freeway segments in California | Loop detectors | 5-min aggregated average speed and volume | Maximize reduction of average of traffic efficiency (product of speed and volume) | 38-47 mph | Estimate lane-specific capacity distributions |
| | Hong et al. (2015) | 2 freeway segments in Illinois and Utah | Loop detectors | 5-min vehicle volume counts, speed, and occupancy | K-means clustering algorithm | NA | Analyze the effect of snow on freeway flow breakdown and recovery |
| | Kidando et al. (2019) | Basic freeway segment in Florida | Microwave vehicle detector | 15-min aggregated average speed | Gaussian mixture model | NA | Evaluate the influence of rainy weather and traffic volume on the dynamic transition of traffic conditions |

2   Note: NA = Non-Applicable; RTMS = remote traffic microwave sensors

A breakdown event was recorded once operating speeds are below these two criteria. The breakpoint speeds were computed for estimating the impact of traffic characteristics and geometric features, such as the presence of high occupancy lanes and ramp meters, on merge-ramps capacity distribution.

To understand the bottlenecks on freeways where lane usage preferences significantly differ, Ma et al. (2012) proposed a lane-based approach to identify a breakdown at divergent sections. In this study, a breakdown phenomenon was visually observed to occur when the speed of the subject lane measured at the detector was lower than its critical speed value. Another condition considered is a sustained speed drop for over 15min to guarantee that the queue propagated. The breakpoint, critical speed in this case, was estimated through optimizing the most significant speed reductions accompanied with breakdown occurrences.

Dehman and Drakopoulos (2012) defined breakdown as a phenomenon that occurred when traffic speed dropped below a critical free-flow speed for at least 15min while the speed at the downstream detector station remained at, or over, this critical speed. As with most of the previous studies that visually identified traffic breakdown, this study identified the critical speed using the fundamental speed-flow diagram. The breakpoint between 50 mph and 60 mph was identified for modeling breakdown process. Later, Dehman (2013) applied a similar approach to identify the breakpoint for capacity distribution modeling. Another study by Laflamme and Ossenbruggen (2014) adopted a visual inspection to define the breakpoint that separates congested and freely flowing traffic regimes.

Filipovska (2019) used a similar method to investigate the effect of weather conditions on the probability of traffic breakdown. The speed drop from a free-mean speed was adopted to define the speed breakpoint. The modified Greenshields' model was used to estimate the free-mean speed from field data. Another study used double-loop detector data to evaluate freeway capacity variations at bottlenecks. The speed breakpoints in the analysis were estimated using a visual inspection of the speed-flow curve. The study adopted breakpoints that ranged from 24.9 mph to 43.5 mph. Other studies (Kim et al. 2010; Dong and Mahmassani 2009) applied a similar approach to identify the beginning of the breakdown event. They found a 10 mph speed drop from a free-flow speed is the appropriate threshold in their analyses. Shao and Liu (2015) adopted 25 mph as breakpoint, a value that was estimated basen on the visual inspection for capacity analysis.

Laflamme and Ossenbruggen (2017) also explored the speed-volume relationship and found that the breakpoint is 50 mph. A similar value of 50 mph was selected by Yeon et al. (2009), who exploited the time series plot of a speed variable to identify the breakpoint for capacity distribution modeling. Despite being the most common method used to estimate the breakpoint, it is prone to producing biased estimates due to high variability in the data. In fact, the traffic fundamental diagram is stochastic and is influenced by many external factors including heterogeneous vehicles, driver behavior, weather conditions, and the random characteristics of demand (Qu et al. 2015; Sopasakis 2004; Chen et al. 2015; Mahnke and Kaupuzs 1999; Xiaobo et al. 2017; Muralidharan et al. 2011; Jabari et al. 2014).

Other studies (Yeon et al. 2009; Kondyli et al. 2013; Shiomi et al. 2011 Shiomi 2016) applied a pre-defined breakpoint for detecting the breakpoint value. A study by Kondyli et al. (2013) developed three approaches to identify the breakdown event: speed, occupancy, and volume-occupancy correlation approaches. The speed or occupancy approach utilized the difference between the two consecutive readings while volume-occupancy correlation approach used the rolling correlation as the breakdown detection criterion. The breakdown in these approaches was identified once the difference or correlation is lower or higher than the breakpoint value, a user-defined parameter. Among the three developed approaches, the speed criterion was recommended as the best performance measure in identifying the breakdown events since it detects breakdown earlier than the other two approaches.

Accordingly, Dowling et al. (2008) applied a somewhat similar approach of using the speed difference between the two consecutive readings to identify the breakdown event. They plotted the speed difference against speed observations and found the speed reading with the highest variability. The study reported 45 mph as the most unstable value in the diagram so it was selected to be the breakdown criterion in the analysis of queue discharge rate.

Unlike the visual observation approach, the use of pre-defined breakpoints may be better because it is not biased by personal judgment. The use of the one-size-fits-all pre-defined breakpoint may provide unreliable results considering that traffic flow is a stochastic process that may vary significantly under different conditions.

Empowered by the advancements of traffic data collection technologies and computational capabilities, some studies have explored the use of data-driven approaches to estimate the breakpoint (Kidando et al. 2019; Xie et al. 2014; Hong, et al. 2015; Kidando et al. 2019). Xie et al. (2014) investigated the freeway capacity heterogeneity among individual lanes using robust statistical estimation methods. In this study, the optimal breakpoint speed was identified for each lane by maximizing the average reduction of traffic efficiency, which was defined as the product of mean speed and volume. The study estimated the lane-specific capacity distributions using a Bayesian hierarchical Weibull model. Hong et al. (2015) used a K-means clustering algorithm for clustering traffic speed data and recognizing typical scenario clusters under the snow effect. The study provided a deeper understanding of the properties of the breakdown and recovery process for a freeway network under the snow effect.

The machine learning algorithms, K-means, and GMM algorithms that have been used in the literature are deterministic in calibrating the traffic flow variables. They also require a large amount of data to yield an optimal solution. This is a well-known problem for the machine learning models, which are prone to overfitting, where a model fails to generalize to data that has not been seen by the model (Kidando et al. 2019).

Despite the significant volume of literature on the stochastic characteristics of traffic breakdown and estimation of the capacity distribution, no study has focused on comparing different statistical models for estimating the breakpoint. Rather, most of the previous research in this area uses either a visual inspection or a pre-defined breakpoint. However, the speed-flow or speed-density/occupancy relationship is highly random and scattered (Kidando et al. 2019; Dehman 2013; Wang, et al. 2009). Therefore it is necessary to develop a robust statistical approach to estimate the breakpoint value that accounts for the data variation and scatteredness in the traffic flow fundamental diagrams. Thus, the main objective of this study is to compare two regression models that can potentially be used to dynamically estimate the breakpoint while also accounting for some specific site data characteristics. Specifically, the LGF and the two-regime model are compared in this study. The analysis is based on freeway data from I-10 in Jacksonville, Florida. The individual lanes' breakpoints are identified by exploring the speed-occupancy relationship.

**DATA AND METHODOLOGY**

Traffic volume, speed, and occupancy data along I-10 in Jacksonville, Florida were used in this study. These data were collected by microwave detectors from two sites near the exit and entrance ramps. Figure 1 (a) depicts one of the study sites (site 1), which has three lanes and is close to an exit ramp. Figure 1 (b) depicts the second study site (site 2), which has five lanes and is close to an entry ramp. The speed limit along the study corridor is 65 mph. As detailed in the figure, lanes 1 and 2 for site 1 and lanes 1 through 3 for site 2 can be assumed to mostly serve the through traffic. Lane 3 for site 1 and lanes 4 and 5 for site 2 serve the exit and entry traffic, respectively. The traffic flow breakpoints may be different because of the differences in the behavior of the traffic served by the two groups of lanes.

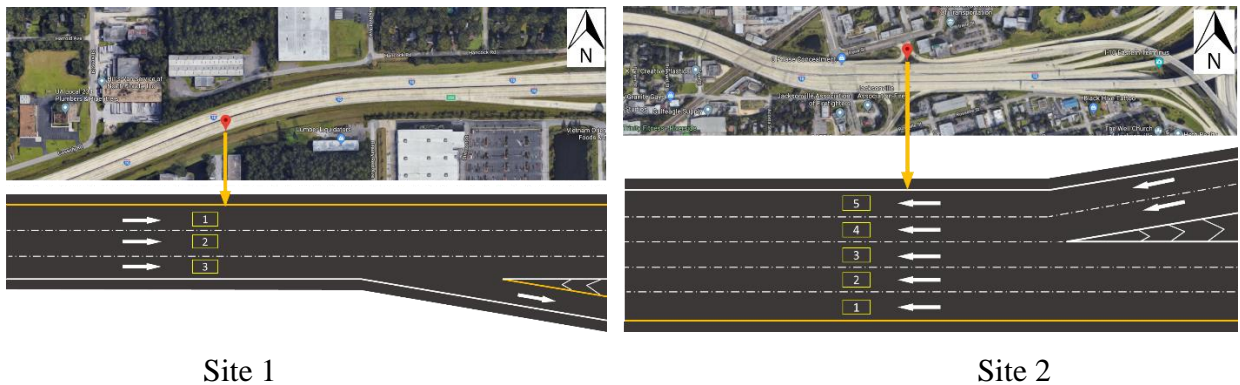

Site 1                                                                                           Site 2

**Fig. 1.** Conceptual diagram of the site characteristics

For modeling purposes, historical traffic data gathered for the period between January 1, 2018, through December 31, 2018 (excluding weekends and holidays) was included in the analysis. Data on Mondays and Fridays were also omitted in the analysis. These data were aggregated at a 5-minute interval. Since the two-hour intervals (i.e. from 6 a.m. to 9 a.m. and 3 p.m. to 7 p.m.) were identified as the busiest hours, these intervals were further evaluated to check if enough data with breakdown events are available for the analysis. Roadways normally experience more breakdowns during peak hours than any other hour in a day (Kidando et al. 2019). Dividing data into intervals for analyzing traffic conditions characteristics is consistent with the previous studies (Qi and Ishak 2014; Guo and Li 2011).

**Modeling methodology**

As indicated earlier, the LGF and two-regime models are compared in detecting breakpoint for breakdown event. In modeling, both regression models explore the relationship between two measured traffic variables, i.e., speed and occupancy. The next sections elaborate the mathematical formulation of the two regression models that were implemented in this study. The two developed models were also compared in terms of their goodness-of-fit using accuracy metrics.

*Two-regime Regression Model*

The two-regime model consists of two regression lines that are separated by a breakpoint (Drake et al. 1967; May 1990). In this study, the breakpoint was treated as an unknown location, calibrated based on the data characteristics by the model. This model is well known as the change-point regression in time series analysis. Assume that the breakpoint value, $\lambda$, separating two expected values ($\mu_{1i}$ and $\mu_{2i}$) and data deviations ($\sigma_1$ and $\sigma_2$). These parameters correspond to the regressions before and after the breakpoint value. Mathematically, this regression model can be expressed as:

$$y_i | \beta_{10}, \beta_{11}, \beta_{20}, \beta_{21}, \lambda, \sigma_1, \sigma_2, x_i \sim \begin{cases} N(\mu_{1i}, \sigma_1^2), & \text{where } x_i \leq \lambda \\ N(\mu_{2i}, \sigma_2^2) & \text{otherwise} \end{cases} \quad (1)$$

where,

$\mu_{1i} = \beta_{10} + \beta_{11} x_i$

$\mu_{2i} = \beta_{20} + \beta_{21} x_i$

$y$ is the observed speed variable,

$x$ represents the observed occupancy variable,

$\beta_{10}, \beta_{11}, \beta_{20}, \text{ and } \beta_{21}$, are the regression coefficients,

$\sigma_1$ and $\sigma_2$ are the standard deviation of the data before and after breakpoint,

$N$ stands for the Gaussian distribution,

*Logistic Function (LGF)*

In this study, the LGF is used as a single-regime model. As formulated in a study by Wang et al. (2011), the LGF with the inflection point, breakpoint value $\lambda$, can be expressed as:

$$y_i | S_{min}, S_{free}, \lambda, \sigma, x_i \sim N(\alpha_i, \sigma^2) \quad (2)$$

$\alpha_i = S_{min} + \frac{S_{free} - S_{min}}{1 + e^{(x_i - \lambda)}}$

where,

$S_{min}$ and $S_{free}$ are the regression coefficients and

$\sigma$ is the standard deviation of the data in the model.

As indicated earlier, the inference and estimation of the two regression models were made using the MCMC simulations. This approach requires specifying the prior density of each parameter. The breakpoint parameter $\lambda$ in the two-regime model was assigned a uniform distribution ($\lambda \sim Uniform(\min_s, \max_s)$). The $\min_s$ and $\max_s$ are the minimum and maximum speed observed in the dataset to allow $\lambda$ to have an equal probability to be at any recorded speed value in the dataset. The prior density of this parameter can be narrowed to reflect the visual inspection of the location of the breakpoint. For the regression parameters ($\beta_{10}, \beta_{11}, \beta_{20},$ and $\beta_{21}$), the prior distributions were assumed to follow the normal distribution with zero mean and variance of 100. Moreover, the standard deviations $\sigma_1$ and $\sigma_2$, were assigned to follow a Half-Normal distribution, $Half-Normal(5)$. For the LGF, $S_{min}$, $S_{free}$, $\lambda$ prior distribution were assigned to follow normal distribution with zero mean and variance of 100. On the other hand, the standard deviation was assigned a similar prior distribution as those assigned in the two-regime regression, $Half-Normal(5)$. All prior distributions assigned in both models are considered to be non-informative priors. The non-informative prior densities allow the likelihood function to dominate the inference process so that the priors have little influence over the estimates (Kruschke 2013).

An open-source Python package based on *PyMC3 3.6* (Salvatier et al. 2016) was used to implement the MCMC simulations. The No-U-Turn Sampler (NUTS) sampling step was applied in which the initial burn-in phases were set to 25,000 iterations and the subsequent 25,000 iterations were used for making inference on the model parameters. A visual diagnostic approach, based on the trace and autocorrelation plots, was used to assess the convergence of the fitted model. Additionally, a Gelman-Rubin diagnostic statistic that estimates the variation with the chains as well as among multiple chains was used. A model is said to have converged when this statistic is one.

*Goodness-of-fit statistics*

The root mean square error (RMSE) of each model is calculated to showcase the performance of the developed models (Equation 3).

$$RMSE = \sqrt{\frac{1}{n}\sum_{i=1}^{n}(y_i - \hat{y}_i)^2} \qquad (3)$$

where,

$y_i$ is the observed speed;

$\hat{y}_i$ is the predicted speed by the model and;

$\bar{n}$ is the total number of the data.

**RESULTS AND DISCUSSION**

Table 2 presents the estimated speed and occupancy breakpoints, with their corresponding 95% Bayesian credible interval (BCI), for the two models fitted in this study (i.e. the LGF and the two-regime model). As one of the single-regime models, the LGF provides only one breakpoint for speed, while the two-regime model produces two speed breakpoints for the same occupancy breakpoint. This is due to the stepwise nature of the multi-regime approach. The region between the lower and the upper speed thresholds defines the transitional state.

As stipulated in Table 2, the occupancy breakpoints estimated by the LGF are consistently lower than those estimated by the two-regime model for site 1. Similarly, the LGF speed breakpoints are overall higher than those of the two-regime model. For the two-regime model, the transitional state that is defined by the two speed thresholds is observed to increase. That is, the magnitude of the differences in the breakpoints kept on increasing from the left-most lane (lane 1) to the right-most lane (lane 3). Since lane 3 tends to serve the exit traffic unlike lanes 1 and 2 of site 1, it has the highest occupancy breakpoints and the lowest speed breakpoints, as expected. Figures 2 (a) through (c) visualize field data and the fitted curves for speed–occupancy relationships. Logically, while lane 1 has the least variation in the traffic, as indicated by the differences in speed 1 and speed 2 breakpoints in Figures 2 (b), (d), and (f). Specifically, the speed differences between lanes 1, 2, and 3, which serve as indicators of the transition state, is 13.91 mph (i.e., 54.95 mph – 41.04 mph), 15.7 mph, and 22.04 mph, respectively.

**Table 2. Estimated breakpoint and their 95 percent credible interval**

| Site and lane number | Breakpoint | LGF | | | Two-regime model | | |
|---|---|---|---|---|---|---|---|
| | | Mean (Std.) | 95% Credible interval | | Mean (Std.) | 95% Credible interval | |
| Site 1-1 | Occupancy (%) | 7.30 (0.29) | 6.76 | 7.86 | 11 | 11 | 11 |
| | Speed 1 (mph) | 55.50 (0.50) | 54.53 | 56.39 | 41.04 | 40.72 | 41.4 |
| | Speed 2 (mph) | na | na | na | 54.95 | 54.61 | 55.26 |
| Site 1-2 | Occupancy (%) | 8.38 (0.33) | 7.73 | 9 | 13.25 | 12 | 17 |
| | Speed 1 (mph) | 60.75 (0.71) | 59.44 | 62.12 | 40.05 | 28.66 | 44.27 |
| | Speed 2 (mph) | na | na | na | 55.75 | 48 | 58.6 |
| Site 1-3 | Occupancy (%) | 11.99 (0.29) | 11.45 | 12.56 | 17 | 17 | 17 |
| | Speed 1 (mph) | 55.13 (0.60) | 54.02 | 56.21 | 29.63 | 29.06 | 30.11 |
| | Speed 2 (mph) | na | na | na | 51.67 | 51.14 | 52.19 |
| Site 2-1 | Occupancy (%) | 11.29 (0.38) | 10.59 | 12.08 | 10 | 10 | 10 |
| | Speed 1 (mph) | 48.74 (0.48) | 47.86 | 49.66 | 47.23 | 46.92 | 47.53 |
| | Speed 2 (mph) | na | na | na | 50.76 | 50.45 | 51.08 |
| Site 2-2 | Occupancy (%) | 19.96 (0.15) | 19.67 | 20.25 | 20.25 | 20 | 21 |
| | Speed 1 (mph) | 44.29 (0.24) | 43.85 | 44.74 | 39.83 | 38.16 | 40.57 |
| | Speed 2 (mph) | na | na | na | 47.07 | 45.49 | 47.79 |
| Site 2-3 | Occupancy (%) | 21.44 (0.18) | 21.1 | 21.77 | 22 | 22 | 22 |
| | Speed 1 (mph) | 43.09 (0.26) | 42.61 | 43.59 | 36.78 | 36.46 | 37.13 |
| | Speed 2 (mph) | na | na | na | 46.6 | 46.39 | 46.81 |
| Site 2-4 | Occupancy (%) | 16.65 (0.30) | 16.08 | 17.23 | 13 | 13 | 13 |
| | Speed 1 (mph) | 46.95 (0.38) | 46.25 | 47.67 | 49.77 | 49.42 | 50.15 |
| | Speed 2 (mph) | na | na | na | 54.27 | 54 | 54.55 |
| Site 2-5 | Occupancy (%) | 6.15 (0.49) | 5.2 | 7.09 | 18.06 | 18 | 19 |
| | Speed 1 (mph) | 55.24 (0.95) | 53.42 | 56.96 | 36.21 | 34.95 | 36.78 |
| | Speed 2 (mph) | | | | 31.56 | 30.49 | 32.71 |

Note: na = not applicable.

As shown in Figure 1, site 2 is a system to system interchange connecting the I-10 and I-95 freeways. Lane 1 is merging I-10 from I-95 southbound and a local road, while lanes 2 and 3 are merging from I-95 southbound traffic from a different exit. Lanes 4 and 5 are joining I-10 from I-95 northbound traffic. The breakpoint estimates from the LGF in Figure 3 suggest that lane 3 (occupancy breakpoint = 21.44%) is more congested than lane 2 (occupancy breakpoint = 19.96%). A similar pattern is observed in the breakpoints estimated using the two-regime model. In contrast, the LGF and two-regime models provided different results. The LGF indicated that lane 4 has a lower occupancy breakpoint than lane 5, while the two-regime model revealed that lane 5 is more congested than lane 4.

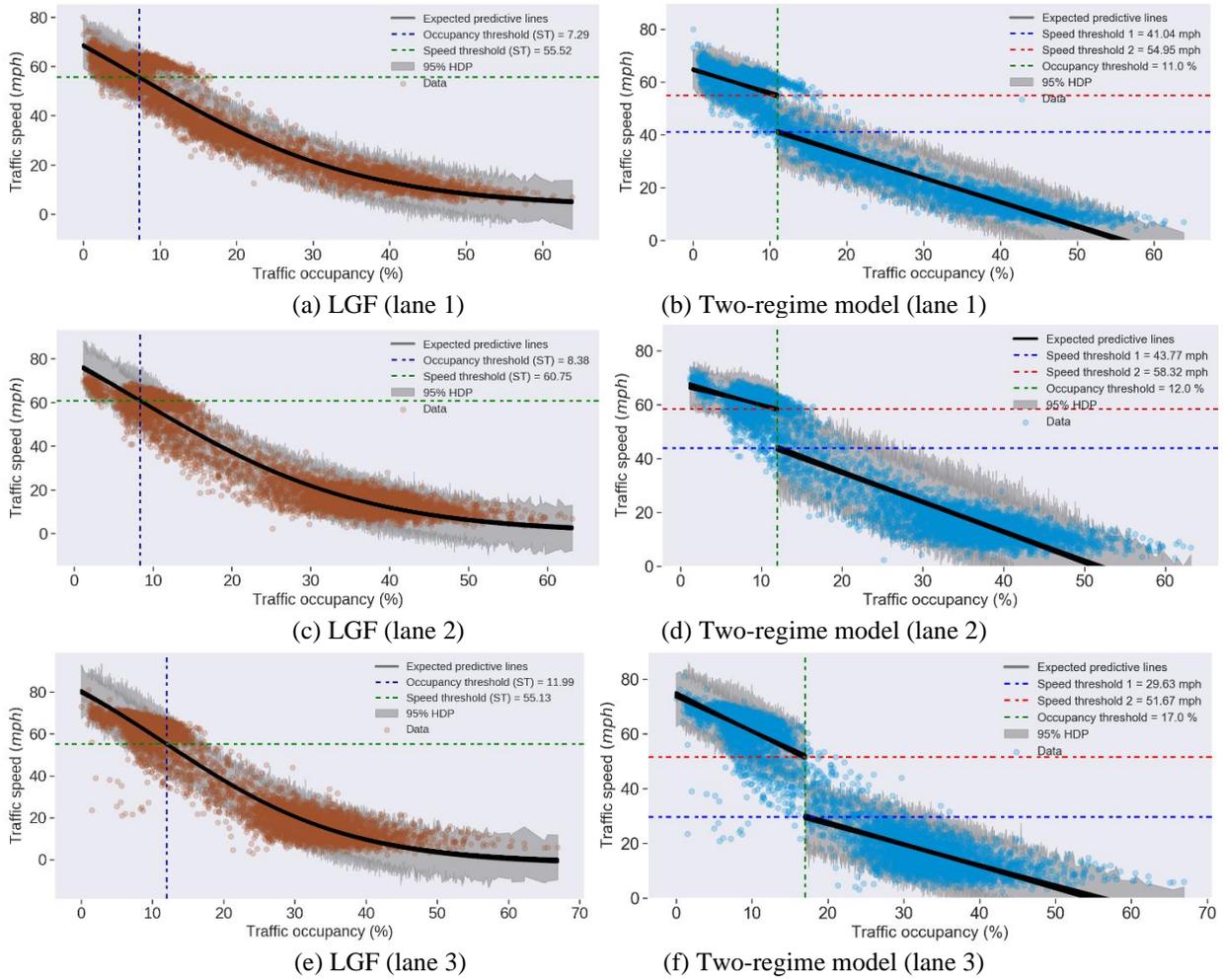

**Fig. 2.** Site 1 field data and fitted curves for speed–occupancy relationships

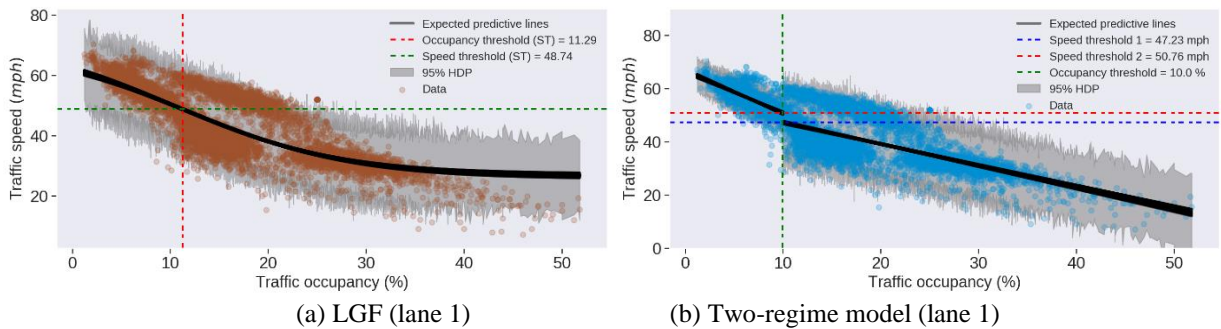

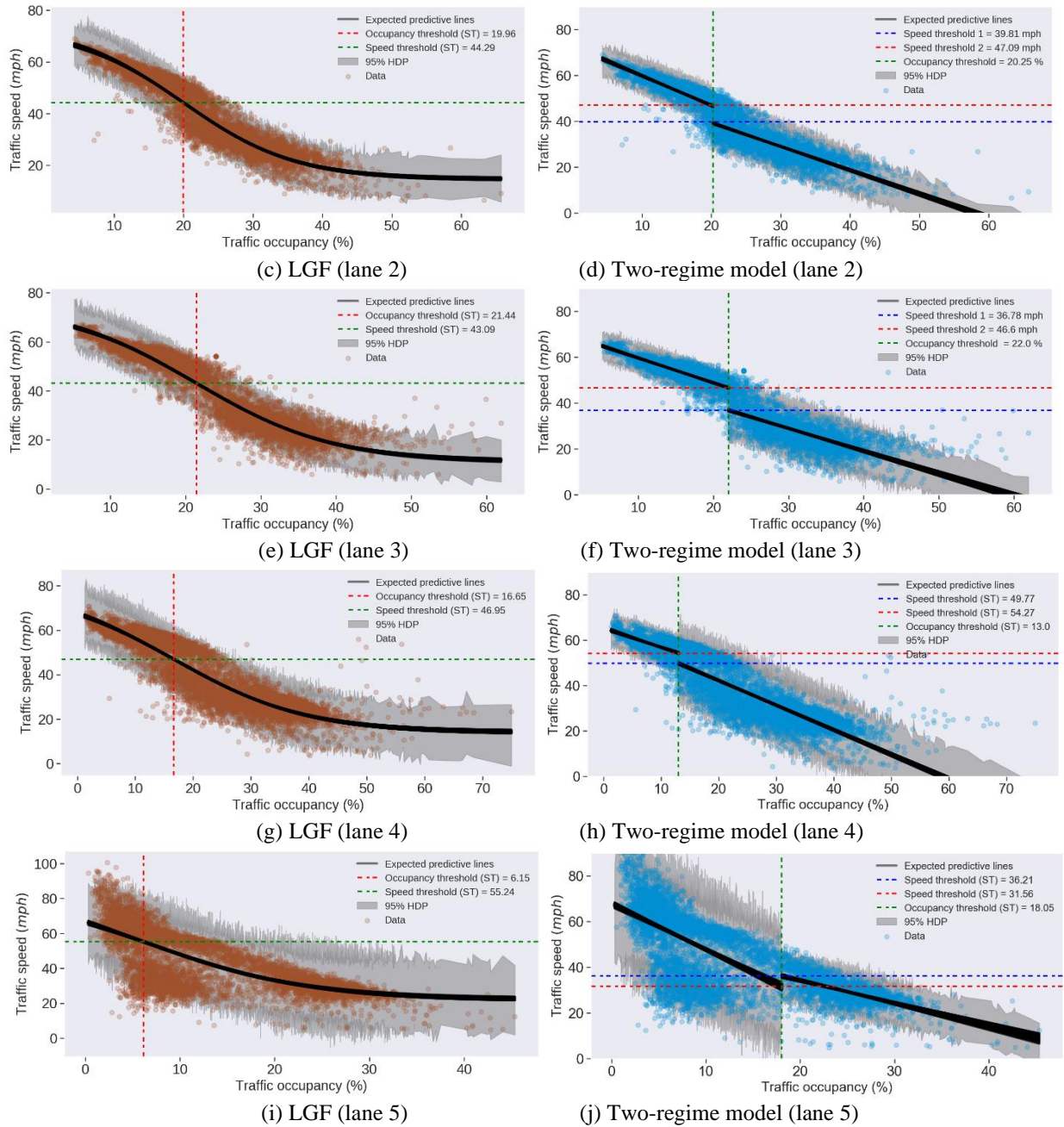

**Fig. 3. Site 2 field data and fitted curves for speed–occupancy relationships.**

One of the advantages of using the proposed estimation in calibrating the speed-occupancy relationship is addressing the stochastic characteristics of this relationship. The proposed calibration provides a distribution with a range of values instead of estimating a point estimate for each parameter in the model. The highest density interval (HDI) is one way to summarize the estimated parameter distribution by providing the most credible values that fall with a certain likelihood. For instance, when the 95% HDI is used for inference it implicitly shows that any value inside the HDI has higher credibility than those outside the interval (Kruschke 2013). Thus, the HDI has the most likely estimates of the model parameter. Figure 4 shows the posterior distribution of the speed breakpoint estimated by the two-regime model. This demonstrates how the proposed analysis accounts for the uncertainty associated with the stochasticity of the speed-occupancy relationship.

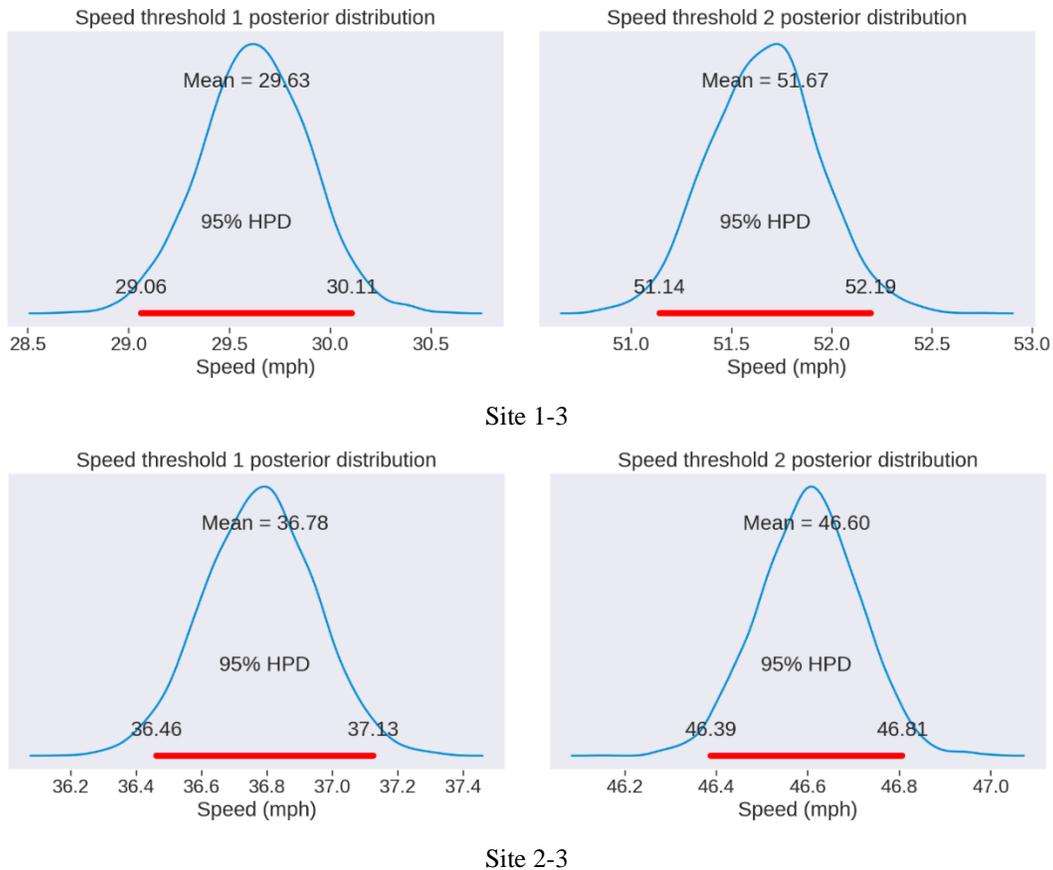

Site 1-3

Site 2-3

**Fig. 4. Posterior distribution of breakpoint for lane 5.**

*Performance of the Models*

Although both the LGF and the two-regime models have shown consistent patterns in estimating the breakpoints, comparing their performances is crucial. As indicated, the performances of these models are compared based on their root mean square error (RMSE). The lower the RMSE, the better the model. Looking at the posterior predicted lines in Figures 2 and 3, the two models performed nearly the same in calibrating the speed-occupancy relationship. Similar results were obtained when the RMSE were estimated using all dataset of each lane. To obtain a better evaluation criterion, the data were divided into portions in the analysis. Figure 5 shows the RMSE for the two models at different ranges of traffic speed. This shows that, in general, the LGF fitted the speed-occupancy relationship with a relatively higher accuracy than the two-regime model, especially in 40-50 mph range. Except lane 1 for site 2, all other lanes for site 1 and 2 were estimated with lower RMSE by the LGF than the two-regime model (see Figure 5). Note that the value in the figure legend shows the average RMSE, while in the parentheses is the standard deviation of the RMSE across the speed range. Consistent with the average estimated RMSE findings, the LGF performed better in the lower and upper traffic speed region than the two-regime model. Five out of eight fitted lane data were estimated with lower RMSE by the LGF than the two-regime model. Moreover, the two-regime model performed relatively poorly in the transitional regime (i.e., at the region near breakpoints), as expected.

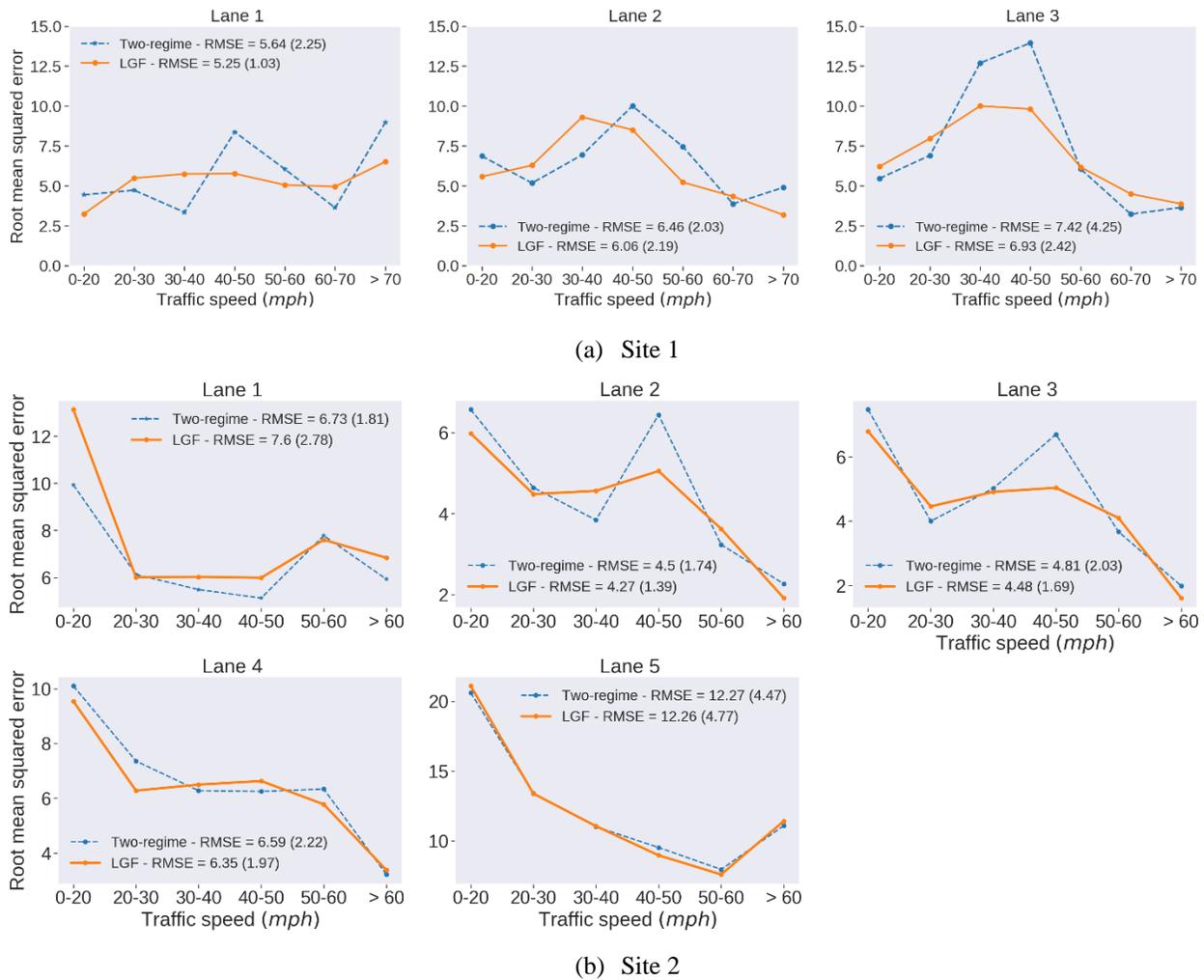

**Fig. 5.** Root mean square error of the LGF and the two-regime models

*Discussion*

The LGF was found to have slightly better accuracy than the two-regime model. Although the two-regime model performed poorer than the LGF, it can be useful in identifying the transitional state. In this case, the range of traffic speed at which the traffic flow regime transitions from congested to uncongested and otherwise can be estimated by the two-regime model. This process could very important in the analysis of breakdown and recovery phenomena. Figure 6 shows the boundaries of the transitional state estimated by the two-regime model for site 1 lane 3 and 2 lane 3.

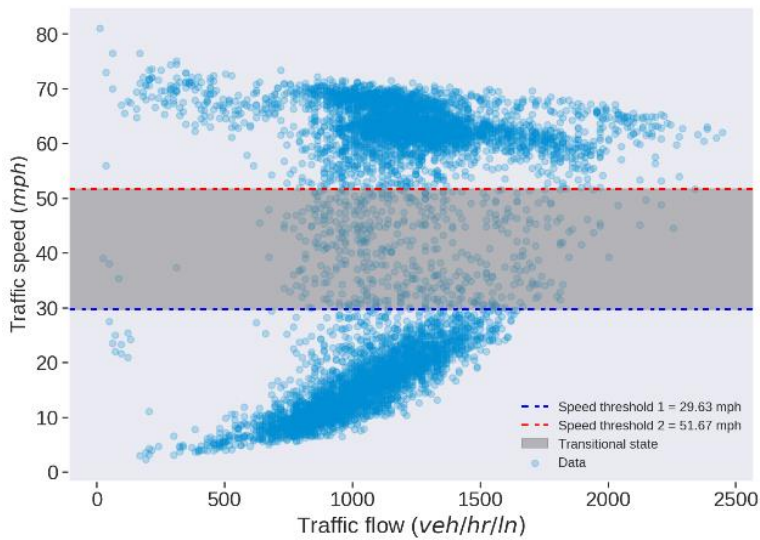

Site 1-3

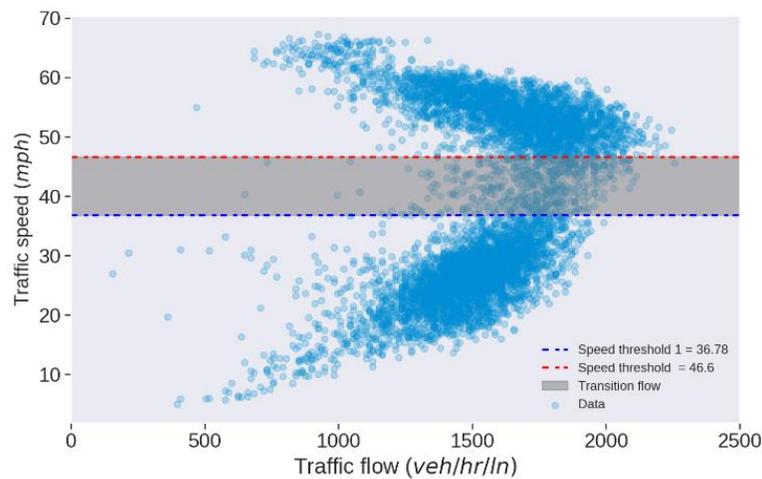

Site 2-3

**Fig. 6.** Estimated speed breakpoints using the two-regime regression on the speed-flow relationship

The LGF was calibrated with only four parameters instead of seven parameters in the two-regime model. Because of this, the computation effort in the LGF is relatively less intensive in terms of computational resources and time than the two-regime model, especially when MCMC estimation approach is considered to account for the stochastic characteristics of the speed-occupancy relationship. The computation time and resources increase with the amount of data used in the calibration process.

**CONCLUSION AND FUTURE WORK**

This study evaluated and compared two regression models – single regime model based on the LGF and the two-regime model – for calibrating the traffic speed-occupancy relationship that can be used to estimate the breakpoint among other applications. The comparison analysis in this research used field data collected along I-10 in Jacksonville, Florida. These data were collected by microwave detectors on two sites (one close to the exit ramp and another close to the merging ramp) for a year (2018).

The analysis has demonstrated that the LGF and the two-regime models showed consistent patterns in estimating the breakpoints. It should be noted that, a consistent pattern of both the occupancy and the speed breakpoints was observed on site that is close to an exit ramp. Contrarily, the breakpoints for the five lanes site (far from the exit ramp) shows a different pattern both as a function of the lane position and the thresholds estimated using the LGF and the two-regime model. The performance of the two models was further compared based on the root mean square error (RMSE). The model performance results indicated that two developed models performed nearly the same in calibrating the speed-occupancy relationship. In terms of RMSE, the LGF generally estimated the speed-occupancy relationship with slightly higher accuracy than the two-regime model especially in the region near the unstable or transitional traffic state.

Since the breakpoints estimate of the two models indicated a similar pattern, the choice of using either the LGF or the two-regime model may depend on the primary objective of the future study. For instance, the two-regime regression model presented estimates of two speed breakpoint values because the model is discontinuous at a breakpoint. Using this model, the region with transitional traffic flow characteristics could clearly be identified. This is one of the benefits of using the two-regime regression model over the LGF, particularly when the transitional flow state is to be explored in the analysis.

On the other hand, the benefit of using the LGF over the two-regime model is its simplicity. The LGF has only four parameters while the two-regime model consists of seven parameters that are to be calibrated using empirical data. In this regard, the two-regime regression model is more computationally intensive than the LGF, especially when the Markov Monte Carlo (MCMC) simulations are adopted in the analysis to account for the random characteristics of the speed-density/occupancy relationship.

A possible extension of this work in the future could be comparing these two parametric models with the Bayesian machine learning algorithms. Examples of machine learning algorithms that could be sought for comparison may include clustering models (e.g. the Bayesian mixture model algorithm), the Hidden Markov Model, the Bayesian Network, etc. The LGF with four parameters was selected in the analysis for this study. Another S-curved function could be used to calibrate the speed-density/occupancy relationship, such as the cumulative standard normal distribution, the Weibull model (two and three-parameters), the Richard function, and Complementary log-log (Gompertz) functions. In other fields of applications, these models are also referred to as the growth models. Furthermore, it would be interesting to compare the S-curved models and other multi-regime models, such as the Edie and the modified Greenberg model in detecting breakpoints. There are also several approaches that can be adopted to improve the performance of the two-regime model, particularly in the region near the breakpoint. An example of such approaches is integrating a non-linear function to allow a smooth transition between the two fitted regimes. This could be an opportunity for future research.